# Graphically E-Learning introduction and its benefits in Virtual Learning

A. Daneshmand Malayeri, *Member IAENG* , J.Abdollahi, R.Rezaei

*Abstract—* **E-learning with using multimedia and graphical interfaces is now fashionable in some virtual learning environments. Especially, in open colleges, universities and E-learning databases, using these interfaces can improve quality of educating by increasing attraction of educational subjects. In this paper, we introduce this technology and its aspects by defining some Graphical User Interfaces (GUI).**

**Improving some indexes in E-learning environments can be measured by using GUI. Adding some plug-ins in E-learning softwares and environments like relative sound, electronic noting paper and virtual classrooms can be created by E-learning GUI (ELGUI) as explain in this paper.**

## I. INTRODUCTION

USING graphically E-learning systems is like making a better and attractive atmosphere for all of virtual education centre.

Most of online students in online colleges and universities believe that using some graphically plug-ins can make a environment for better educational feedback. Analyzing statistics of using GUI in E-learning and E-education show that providing these environments have better influence than simple environments.

Relative Sound Effecting (RSE) and Electronic Noting Paper (ENP) are the most application of using plug-ins in E-learning and E-education systems. Some websites that present E-learning and E-education systems used from these plug-ins in the best way [3].

Generally, one of ways for increasing rate of website visitors will be caused by adding useful and related plug-ins and graphically interfaces, Especially in educating websites.

Today, e-learning systems exist to serve the needs of distant learners and as a supplement to traditional teaching methods.

A.Daneshmand Malayeri and J.Abdollahi are with the Hamedan University of Technology, Computer Engineering Department , Hamedan, Iran (corresponding author to provide phone: 0098-851-2252351; fax: 0098-851-2230154; e-mail: amin.daneshmand@gmail.com ).

A.Daneshmand Malayeri is also with the Young Researchers Club of Malayer Azad University; phone: 0098-851-2230153 ; e-mail : admalayeri@yahoo.com

R.Rezaei is with Malayer Azad University, Malayer, Iran, phone : 0098-851-2228093; e-mail : rezaee@iau-malayer.ac.ir

Many of these systems function mainly as management services for course material and their users [1] .

Communication takes the form of asynchronous chat rooms, forums and message boards, with the learning material being presented mainly in text form. These learning systems have failed to take full advantage of the availability of high-speed Internet connections to de- liver a more intuitive learning environment with many media types, which offer a stimulating way for students to learn, socialize and collaborate. Traditional e-learning systems fail to address has shown that collaborative and group work can assist students in attaining a higher achievement level. The importance of classmates is also recognized as an important factor for succeeding in education [2].

There is much deliberation in the literature concerning instructional methodology and design. Discussion ranges from considerations of media options and technological hurdles to attempts to distinguish between online and traditional class materials .Existing pedagogical research is limited, according to Frydenberg, who suggests that this is because few fully developed programs have arrived at a stage where summative evaluation is possible [4]. Studies concern mainly the application of traditional teaching concepts to e-learning environments. For example, Roslin Brennan cites communication, interactivity, and social cohesion as key pedagogical goals, while George Siemens contends that variety is a central requirement for learning, and that media choices should be made according to desired learning outcomes. Many of these concepts can be traced back to the work of John Dewey , who, in writing about experiential learning, argued that education must engage with and enlarge experience, and that interaction and environments for learning provide a continuing framework for teaching practice. "This process of making meaningful connections is at the core of all learning". E-learning materials often reflect these concepts through the use of simulations and a focus on interactive learning activities. Jasinski explores improvization as a strategy to make online materials more meaningful to learners by providing a better balance between content and process. Jasinski contends that this strategy enables a fast transition from the conceptual to the operational, and that students "learn by playing with rules, not by rules, or to create new rules"[5]. Educational materials that have been effectively designed will facilitate the achievement of desired learning outcomes for students. Effective design of electronic learning materials relies on instructional design processes that reflect the absence of or reduction in face-to-face instruction.

## II. GRAPHICAL USER INTERFACES IN E-LEARNING SYSTEMS

Using a GUI-Based in an E-learning system has been shown in Figure 1. All of essential parts for using a better environment in E-learning systems has been introduced in this figure. We explain these parts and their benefits and needs.

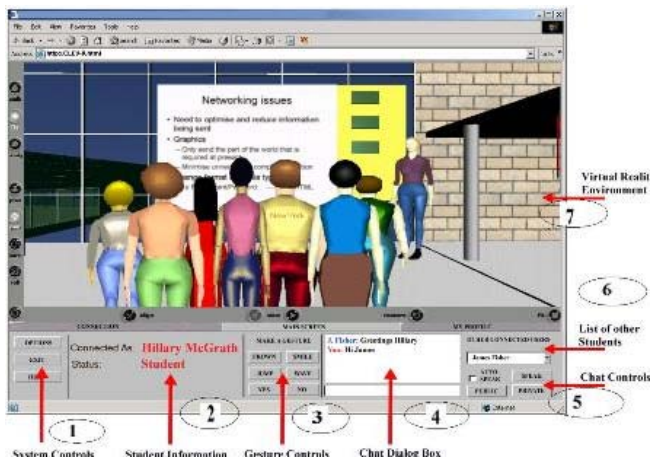

Figure1. GUI using in an E-learning system with its elements

Now, we introduce all of parts from 1-7 in figure 1.

1. System Controls : all of primary and advanced settings for users has been defined in this tab. Users can be able to use it for entrance, exit and option accessories.

2. Student Information : all of students have an unique and secure information for better introducing. Some essential fields for this part is: name, student ID, username and hidden password. One of the most important is for passwords. All of passwords should be out of reach for other students.

3. Gesture Controls: It is a plug-in for make a likely gestures for students. By some GUI, we can add a graphically and attractive environment to an E-learning website in virtual learning.

4. Chat Dialog Box: By this plug-in, a semantic environment has been created. We can define two level chat managing. The first level is between students and the second one is between students and educational consultants. These consultants must be online every educational time. At break times, students can leave offline messages for next reply from educational consultants. All of students can share their knowledge together by this part. This part play an effective role in increasing educational feedback.

5. Chat Controls : This part is using for chat settings. Students can change their kind of chatting include public chat, writing chat and speak chat .

6. List of Other Students: Students can browse other students from this box. All of students that are logged in to website has been shown in this part.

7. Virtual Reality Environment: Graphical experts are the main role in this part. They should design an environment with flexibility in all of E-learning websites and virtual spaces.

## III. REVIEW OF IMPACTS IN GRAPHICALLY AND MULTIMEDIA E-LEARNING ENVIRONMENT

The difference that the e-learning materials will make is an important design consideration. The influence of the e-learning design can be assessed from a number of perspectives, including the way that it will affect the learner, the ramifications that it will have for the learning (and broader) community into which it will be implemented, and the environmental influence of its development and use. Considerations about the personal influence of the e-learning design might consider the extent of learning that is likely to take place compared to the effort required to produce the resource. They might also consider the potential effect of the content and its presentation on a person's self-esteem and other psychological states – in short, the extent to which the content benefits the user. Considerations about the social influence of the e-learning design might include the cultural appropriateness of the material, the extent to which the design makes demands on others working with or supervising the learner, the way that it may influence cultural capital in the educational setting, and the ethical values implicit in the design or content – in short, the extent to which people other than the learner will benefit. Considerations about the environmental influence of the design include the use of resources required to develop and deliver the e-learning materials, and the influence on the environment of activities required by people using or administering the learning materials – in short, the extent to which the environment will benefit from the design. As concludes, "sensibilities regarding people and nature seem central to what technology ought to be about". Considering the influence of the design requires designers to appreciate their relationship with, and influence upon, the learner and the learner's social and physical context. Designers need to act in a responsible and ethical manner to ensure that the impact of their e- learning design is of benefit to the learner, society, and the environment[ 4-6].

## IV. CONTEXT OF GRAPHICALLY AND MULTIMEDIA E-LEARNING ENVIRONMENT

The situation within which the e-learning resources are to be used has a significant influence on the design, but may only be partially predictable. Some aspects of the usage context are implied by the elements previously described; however, the broader context is also relevant to addressing learning needs.

Elements of activity, scenario, and feedback need to take into account the users' profiles and the delivery element needs to consider the technical infrastructure. However, additional contextual considerations include the institutional objectives of the e-learning program, the role and skills of any instructor, longevity of the resources, and cultural sensitivities. The connection between context and delivery methods is highlighted by Silverman and Casazza, who note that "different systems of communication seem to be at the heart of many of the cultural and ethnic differences that affect the learning environment". Bearing this in mind, it is clear that the broader context within which the learning activity is delivered can influence many elements of e-learning design [5,6].

## V. GRAPHICAL AND CLEV-R SYSTEM FUNCTION IN E-LEARNING

The CLEV-R system is a web-based multi-user VR environment that aims to support interactive e-learning through the use of several different multi-media types. In addition to providing a visual interface for the users, features such as video, audio and images enhance the learning experience for the user. The 3D environment of the system is implemented using the VRML ISO standard [7]. This scripting language provides a means of designing a virtual world which can then be rendered and displayed in any VRML enabled Web browser. The CLEV-R project makes use of the freely avail- able Cortona VRML viewer by Parallel graphics [8]. Using VRML it is possible both to design the virtual build- ings where the learning is to take place and to create avatars that will represent the users of the system in the virtual world. Functionality for animation and interactive features are also available so it is possible to define movement for the avatars and multiple different ways for the users to inter- act with the system.

A key factor in the development of CLEV-R was the provision of an intuitive Graphical User Interface (GUI), which is simple to use and comprehend. The GUI for CLEV-R can be seen in figure 1. The GUI consists of two main panels; the upper panel is used for displaying the VRML multi-user environment while the lower panel provides access for users to interact with the system. Three tabbed panels on the lower portion of the screen provide different functionalities for the users. The first panel allows new users to register with the system and authenticates returned users. Users can also se lect their avatars and choose a user name.

One of the main purposes of the GUI is to allow commu- nication among connected users. There are three main ways in which students interact, namely gestures, text and voice. The learner can choose gestures for their avatar to depict in the VR world. For example, users can instruct their avatar to raise their hand and so inform other users they have a question. The learner can enter into real-time text conversations by typing messages into the appropriate textfield on the GUI. A record of all messages is also maintained. Text messages can be sent to certain groups or individuals which can be selected using the drop down menu, showing all con- nected users. Voice conversations can also be controlled via the communication panel; again buttons enable different op- tions to be selected.

## VI. FEEDBACK REVIEWS OF USING GUI AND MULTIMEDIA IN E-LEARNING ENVIRONMENTS

Experience becomes knowledge through reflection, which is enhanced by timely and appropriate criticism. Effective e-learning design will include provision for feedback that amplifies the learning from the experience, and enables students to increase their level of skill and knowledge. The range of available feedback strategies is vast, including reflective responses to prescribed questions, semi-automated responses by the system to student actions and work, shared comments in online forums and blogs, and personal responses via email, telephone, and post. The technologically mediated nature of e-learning is perhaps most apparent in the element of feedback, and the challenges are significant for e-learning in domains that have traditionally relied heavily on interpersonal communication, in particular, psychological counseling and the performing arts where "one-to-one is the traditional norm"[6]. Timeliness of feedback is also a consideration. Timeliness may be enhanced through automation in some cases, or it might be delayed, such as where email responses replace tutorial question and answer sessions. Effective use of feedback will enable an e-learning design to set up a dialogue within which the student participates, without which designs may simply become plans for broadcasting content. In the TLF learning object Sonic Space City, where students create a soundscape to visual stimulus, feedback mechanisms include the ability to monitor work in progress in real time, and the facility to analyze and enhance projects by recording and replaying audio. In addition, students can describe their finished soundscape according to whether they consider it peaceful or noisy, calm or busy, scary or safe. These descriptions may be printed and the views of other class members sought. In this design, deliberate attention has been paid to incorporating multiple avenues for feedback to the learner, both during and after the activity. Students who use this learning object have displayed high levels of motivation and engagement because of the immediate and rich feedback provided by the activity. Activities such as this might be enhanced by enabling feedback from a broader range of people via the presentation of their soundscapes in an online forum, where reviews and downloads could provide qualitative and quantitative feedback about the appeal of their musical creation to others [6].

## VII. CONCLUSION

In this paper, we introduce some methods for increasing attraction of E-learning environments by adding graphically and multimedia interfaces. GUI and CLEV-R are the most applications for using these environments.

According to some statistics and experiences, using GUI, multimedia and CLEV-R technologies in E-learning systems can improve and develop effects of virtual learning and its feedback.

CLEV-R, a multi-user environment for education offers a new way for university students to learn, collaborate and socialize. The use of VR environment in education is a relatively new idea that is gaining interest, fuelled by the advent of broadband technologies and improved Internet connections. CLEV-R takes advantage of enhanced Internet access to provide stimulating learning environments augmented with multi-media. These environments, mimicking a real university setting, delivers all available facilities to the students. This simulation of a real university provides an intuitive means for students and teachers to interact with CLEV- R. A mix of traditional and novel approaches are utilized to ensure network traffic is minimized and the system operates efficiently. Content is filtered depending on individual user's c onnection speeds and efforts are currently under way to make CLEV-R available on mobile devices.

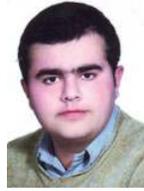

**Amin Daneshmand Malayeri** is now student of Computer Engineering in Hamedan University of Technology, Hamedan , Iran. He is also member of Young Researchers Club of Malayer Azad University and International Association of Engineering (IAENG) . He is now local ambassador of User Experience Network of America in Iran and connector member of Informatics Society of Iran as partner of members committee. His main research interests are ERP and CRM methodologies, adaptive E-Learning systems and Knowledge Management systems.

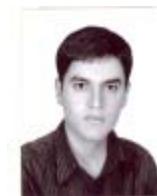

**Jalal Abdollahi** is now student of Computer Engineering in Hamedan University of Technology, Hamedan, Iran.  He is a researcher in ICT technologies and Computer Networks  .His main research interests are in E-Learning and E-Business.

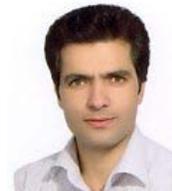

**Rasool Rezaei** is now candidate of M.Sc in Computer Engineering, Hardware architecture in Arak Azad University, Arak, Iran. He received his B.Sc in Software Engineering from Arak Azad University. He is now manager of network and website office of Malayer Azad University. His main research interests are in Network Engineering and software development.